\documentclass{emulateapj}
\usepackage{color}

\def\rlow{ \ifmmode r_{54} \else $r_{54}$\fi}
\def\rmed{ \ifmmode r_{42} \else $r_{42}$\fi}
\def\rhigh{\ifmmode r_{21} \else $r_{21}$\fi}

\def\rall{\ifmmode r_{51}  \else $r_{51}$\fi}

\def\rgrav{\ifmmode R_\mathrm{g}  \else $R_\mathrm{g}$\fi}
\def\rsp{\ifmmode R_\mathrm{sp}  \else $R_\mathrm{sp}$\fi}
\def\littlersp{\ifmmode r_\mathrm{sp}  \else $r_\mathrm{sp}$\fi}
\def\tsp{\ifmmode T_\mathrm{sp}  \else $T_\mathrm{sp}$\fi}
\def\rph{\ifmmode R_\mathrm{ph}  \else $R_\mathrm{ph}$\fi}
\def\littlerph{\ifmmode r_\mathrm{ph}  \else $r_\mathrm{ph}$\fi}
\def\tph{\ifmmode T_\mathrm{ph}  \else $T_\mathrm{ph}$\fi}
\def\tapp{\ifmmode T_\mathrm{app}  \else $T_\mathrm{app}$\fi}

\def\lam{\ifmmode {\lambda} \else {$\lambda$}\fi}

\def\cosi{\ifmmode {\cos\,i} \else $\cos\,i$\fi}

\def\teff{\ifmmode {T_{\mathrm{eff}}} \else $T_{\mathrm{eff}}$\fi}
\def\tdisk{\ifmmode {T_{\rm disk}} \else $T_{\rm disk}$\fi}
\def\tmax{\ifmmode {T_{\rm max}} \else $T_{\rm max}$\fi}
\def\logtmax{\ifmmode {\mathrm{log}\,T_{\rm max}} \else $\mathrm{log}\,T_{\rm max}$\fi}
\def\loglbyled{\ifmmode {\mathrm{log}\,L/\led} \else $\mathrm{log}\, L/\led$\fi}
\def\dlogLR{\ifmmode {d\mathrm{d\, ln}\,L/\mathrm{d\, ln}\,R} \else $\mathrm{d\, ln}\,L/\mathrm{d\, ln}\,R$\fi}
\def\mbh{\ifmmode {M_{\rm BH}} \else $M_{\rm BH}$\fi}
\def\logmbh{\ifmmode {\mathrm{log}}~M_{\mathrm{BH}} \else ~log~$M_{\mathrm{BH}}$\fi}
\def\mdot{\ifmmode {\dot M} \else $\dot M$\fi}
\def\littlemdot{\ifmmode {\dot m} \else $\dot m$\fi}
\def\mdoto{\ifmmode {\dot{M}_0} \else  $\dot{M}_0$\fi}
\def\mdoted{\ifmmode {\dot{M}_{\mathrm{Ed}}} \else  $\dot{M}_{\mathrm{Ed}}$\fi}
\def\lamLlam{\ifmmode \lambda L_{\lambda}(5100) \else {$\lambda
    L_{\lambda}(5100)$} \fi}
\def\llam{\ifmmode {L_\lambda} \else  $L_\lambda$ \fi}
\def\flam{\ifmmode {F_\lambda} \else  $F_\lambda$ \fi}
\def\sigstar{\ifmmode \sigma_* \else $\sigma_*$\fi}
\def\hbeta{\ifmmode {\rm H}\beta \else H$\beta$\fi}
\def\civ{\ifmmode {\rm C{\sc iv}} \else C~{\sc iv}\fi}

\def\ergps{\ifmmode \rm erg~s^{-1} \else $\rm erg~s^{-1}$ \fi}
\def\kmps{\ifmmode \rm km~s^{-1}\else $\rm km~s^{-1}$\fi}
\def\kms{km~s$^{-1}$}
\def\lbol{\ifmmode {L_\mathrm {bol}} \else $L_\mathrm {bol}$\fi}
\def\lnu{\ifmmode {L_\mathrm {\nu}} \else $L_\mathrm {\nu}$\fi}
\def\led{\ifmmode {L_\mathrm {Ed}} \else $L_\mathrm {Ed}$\fi}

\def\mgii{\ifmmode {\rm Mg{\sc ii}} \else Mg~{\sc ii}\fi}

\def\gsim{\lower 2pt \hbox{$\, \buildrel {\scriptstyle >}\over
{\scriptstyle \sim}\,$}}
\def\lsim{\lower 2pt \hbox{$\, \buildrel {\scriptstyle <}\over
{\scriptstyle \sim}\,$}}

\def\simlt{\lower.5ex\hbox{$\; \buildrel < \over \sim \;$}}
\def\simgt{\lower.5ex\hbox{$\; \buildrel > \over \sim \;$}}

\def\gsim{\lower 2pt \hbox{$\, \buildrel {\scriptstyle >}\over
{\scriptstyle \sim}\,$}}
\def\lsim{\lower 2pt \hbox{$\, \buildrel {\scriptstyle <}\over
{\scriptstyle \sim}\,$}}

\def\deg{\ifmmode ^{\circ}
         \else $^{\circ}$\fi}
\def\pdeg{\ifmmode
           $\setbox0=\hbox{$^{\circ}$}\rlap{\hskip.11\wd0 .}$^{\circ}
     \else \setbox0=\hbox{$^{\circ}$}\rlap{\hskip.11\wd0 .}$^{\circ}$\fi}
     
\def\pc{\ifmmode \mathrm{pc} \else $\mathrm{pc}$ \fi}
\def\mpc{\ifmmode \mathrm{Mpc} \else $\mathrm{Mpc}$\fi}
\def\mpcthree{\ifmmode \mathrm{Mpc}^{-3} \else $\mathrm{Mpc}^{-3}$\fi}
\def\gpcthree{\ifmmode \mathrm{Gpc}^{-3} \else $\mathrm{Gpc}^{-3}$\fi}

\def\kelvin{\ifmmode \mathrm{K} \else {$\mathrm{K}$}\fi}
\def\kev{\ifmmode \mathrm{keV} \else $\mathrm{keV}$ \fi}

\def\lsun{\ifmmode {L_\odot} \else $L_\odot$\fi}
\def\msun{\ifmmode M_\odot \else $M_\odot$\fi}
\def\msunyr{\ifmmode M_\odot~\mathrm{yr}^{-1} \else $M_\odot~\mathrm{yr}^{-1}$\fi}

\def\cosi{\ifmmode {\cos\,i} \else $\cos\,i$\fi}

\def\hzero{\ifmmode {\rm H}^0 \else H$^0$\fi}
\def\hplus{\ifmmode {\mathrm{H}}^+ \else H$^+$\fi}
\def\hezero{\ifmmode {\mathrm{He}}^0 \else He$^0$\fi}
\def\heplus{\ifmmode {\mathrm{He}}^+ \else He$^+$\fi}
\def\heplustwo{\ifmmode {\mathrm{He}}^{+2} \else He$^{+2}$\fi}
\def\neplustwo{\ifmmode {\mathrm{Ne}}^{+2} \else Ne$^{+2}$\fi}
\def\neplusfour{\ifmmode {\mathrm{Ne}}^{+4} \else Ne$^{+4}$\fi}

\def\heii{\ifmmode {\rm He~{II}} \else He~{\sc ii}\fi}
\def\civ{\ifmmode {\rm C{IV}} \else C~{\sc iv}\fi}

\def\oii{\ifmmode {\rm O{~II}} \else O~{\sc ii}\fi}
\def\oiii{\ifmmode {\rm O{~II}} \else O~{\sc iii}\fi}

\newcommand{\nev}{{[Ne~{\sc v}]}}
\def\neiii{\ifmmode {\rm Ne~{III}} \else Ne~{\sc iii}\fi}
\def\nev{\ifmmode {\rm Ne~{V}} \else Ne~{\sc v}\fi}
\def\mgii{\ifmmode {\rm Mg{\sc ii}} \else Mg~{\sc ii}\fi}

\def\caii{\ifmmode {\rm Ca~{V}} \else Ca~{\sc ii}\fi}
\newcommand{\feii}{Fe~{\sc ii}}

\def\mbh{\ifmmode {M_{\rm BH}} \else $M_{\rm BH}$\fi}
\def\led{\ifmmode L_{\mathrm{Ed}} \else $L_{\mathrm{Ed}}$\fi}
\def\lbolflare{\ifmmode L_{\mathrm{bol,flare}} \else $L_{\mathrm{bol,flare}}$\fi}
\def\lagn{\ifmmode L_{\mathrm{agn}} \else $L_{\mathrm{agn}}$\fi}
\def\lbolagn{\ifmmode L_{\mathrm{bol,agn}} \else $L_{\mathrm{bol,agn}}$\fi}
\def\lbol{\ifmmode L_{\mathrm{bol}} \else $L_{\mathrm{bol}}$\fi}
\def\mdot{\ifmmode {\dot M} \else $\dot M$\fi}
\def\mdoto{\ifmmode {\dot{M}_0} \else  $\dot{M}_0$\fi}
\def\mdotf{\ifmmode {\dot{M}_\mathrm{flare}} \else  $\dot{M}_\mathrm{flare}$\fi}

\def\hnot{\ifmmode H_0 \else H$_0$ \fi}

\def\vkep{\ifmmode v_\mathrm{Kep} \else $v_\mathrm{Kep}$ \fi}
\def\vc{\ifmmode v_\mathrm{c} \else $v_\mathrm{c}$ \fi}

\def\vthree{\ifmmode v_{1000} \else $v_{1000}$ \fi}
\def\vrel{\ifmmode v_\mathrm{rel} \else $v_\mathrm{rel}$ \fi}
\def\vkick{\ifmmode v_\mathrm{kick} \else $v_\mathrm{kick}$ \fi}
\def\vkickz{\ifmmode v_{\mathrm{kick},z} \else $v_{\mathrm{kick},z} $ \fi}
\def\vkicky{\ifmmode v_{\mathrm{kick},y} \else $v_{\mathrm{kick},y} $ \fi}
\def\vchar{\ifmmode v_\mathrm{char} \else $v_\mathrm{char}$ \fi}
\def\eflare{\ifmmode E_\mathrm{flare} \else $E_\mathrm{flare}$ \fi}
\def\ekick{\ifmmode E_\mathrm{kick} \else $E_\mathrm{kick}$ \fi}
\def\ecoll{\ifmmode E_\mathrm{coll} \else $E_\mathrm{coll}$ \fi}
\def\ezero{\ifmmode E_\mathrm{0} \else $E_\mathrm{0}$ \fi}
\def\efac{\ifmmode \xi_\mathrm{E} \else $\xi_\mathrm{E}$ \fi}
\def\tqso{\ifmmode t_\mathrm{QSO} \else $t_\mathrm{QSO}$ \fi}
\def\tflare{\ifmmode t_\mathrm{flare} \else $t_\mathrm{flare}$ \fi}
\def\tzero{\ifmmode t_\mathrm{0} \else $t_\mathrm{0}$ \fi}
\def\tfac{\ifmmode \xi_\mathrm{t} \else $\xi_\mathrm{t}$ \fi}
\def\gfac{\ifmmode f_\mathrm{g} \else $f_\mathrm{g}$ \fi}
\def\lflare{\ifmmode L_\mathrm{flare} \else $L_\mathrm{flare}$ \fi}
\def\fflare{\ifmmode F_\mathrm{flare} \else $F_\mathrm{flare}$ \fi}
\def\nflare{\ifmmode N_\mathrm{flare} \else $N_\mathrm{flare}$ \fi}
\def\tshock{\ifmmode T_\mathrm{shock} \else $T_\mathrm{shock}$ \fi}
\def\rmin{\ifmmode R_\mathrm{1} \else $R_\mathrm{1}$ \fi}
\def\rmax{\ifmmode R_\mathrm{2} \else $R_\mathrm{2}$ \fi}
\def\rbound{\ifmmode R_\mathrm{b} \else $R_\mathrm{b}$ \fi}
\def\pbound{\ifmmode P_\mathrm{b} \else $P_\mathrm{b}$ \fi}
\def\mbound{\ifmmode M_\mathrm{b} \else $M_\mathrm{b}$ \fi}
\def\mbo{\ifmmode M_{\mathrm{b}0} \else $M_{\mathrm{b}0} $ \fi}
\def\ebo{\ifmmode E_{\mathrm{b}0} \else $E_{\mathrm{b}0} $ \fi}
\def\efinal{\ifmmode E_\mathrm{final} \else $E_\mathrm{final} $ \fi}
\def\tbound{\ifmmode t_\mathrm{b} \else $t_\mathrm{b}$ \fi}
\def\tagn{\ifmmode t_\mathrm{AGN} \else $t_\mathrm{AGN}$ \fi}
\def\torb{\ifmmode t_\mathrm{orb} \else $t_\mathrm{orb}$ \fi}
\def\tdf{\ifmmode t_\mathrm{df} \else $t_\mathrm{df}$ \fi}
\def\rlim{\ifmmode R_\mathrm{lim} \else $R_\mathrm{lim}$ \fi}
\def\vlim{\ifmmode v_\mathrm{lim} \else $v_\mathrm{lim}$ \fi}
\def\vphi{\ifmmode v_\phi \else $v_\phi$ \fi}
\def\mlim{\ifmmode M_\mathrm{lim} \else $M_\mathrm{lim}$ \fi}
\def\tlim{\ifmmode t_\mathrm{lim} \else $t_\mathrm{lim}$ \fi}
\def\llim{\ifmmode L_\mathrm{lim} \else $L_\mathrm{lim}$ \fi}
\def\fqso{\ifmmode f_\mathrm{QSO} \else $f_\mathrm{QSO}$ \fi}
\def\phii{\ifmmode \phi_{\mathrm i}\else $ \phi_{\mathrm i}$\fi}

\def\alphaos{\ifmmode \alpha_{os} \else $\alpha_{os}$\fi}

\def\hbeta{\ifmmode \rm{H}\beta \else H$\beta$\fi}
\def\hbetan{\ifmmode \rm{H}\beta_{\rm n} \else H$\beta_{\rm n}$\fi}
\def\hgamma{\ifmmode \rm{H}\gamma \else H$\gamma$\fi}
\def\hdelta{\ifmmode \rm{H}\delta \else H$\delta$\fi}
\def\hepsilon{\ifmmode \rm{H}\epsilon \else H$\epsilon$\fi}
\def\hzeta{\ifmmode \rm{H}\zeta \else H$\zeta$\fi}
\def\halpha{\ifmmode \rm{H}\alpha \else H$\alpha$\fi}
\def\lalpha{\ifmmode \rm{Ly}\alpha \else Ly$\alpha$\fi}

\def\dvhb{\ifmmode \Delta v_{\hbeta} \else $\Delta v_{\hbeta}$\fi}
\def\dvmg{\ifmmode \Delta v_{\rm{Mg}} \else $\Delta v_{\rm{Mg}}$\fi}

\def\muobs{\ifmmode {\mu_{o}} \else  $\mu_{o}$ \fi}
\def\cosi{\ifmmode {\mathrm{cos}\,i} \else $\mathrm{cos}\,i$\fi}

\def\tauh{\ifmmode {\tau_{\rm H}} \else $\tau_{\rm H}$ \fi}

\def\yr{\ifmmode {\rm yr} \else  yr \fi}
\def\kms{\ifmmode \rm km~s^{-1}\else $\rm km~s^{-1}$\fi}
\def\cm{\ifmmode {\rm cm} \else  cm \fi}
\def\cmmitwo{\ifmmode \rm cm^{-2} \else $\rm cm^{-2}$\fi}
\def\cmmitwops{\ifmmode \rm cm^{-2}~s^{-1} \else $\rm cm^{-2}~s^{-1}$\fi}
\def\cmmithree{\ifmmode \rm cm^{-3} \else $\rm cm^{-3}$\fi}
\def\cmps{\ifmmode \rm cm~s^{-1}\else $\rm cm~s^{-1}$\fi}
\def\cmpsps{\ifmmode \rm cm~s^{-2}\else $\rm cm~s^{-2}$\fi}
\def\kmps{\ifmmode \rm km~s^{-1}\else $\rm km~s^{-1}$\fi}
\def\kmpspmpc{\ifmmode \rm km~s^{-1}~Mpc^{-1} \else
    $\rm km~s^{-1}~Mpc^{-1}$\fi}
  
\def\gcmthree{\ifmmode \rm g~cm^{-3} \else $\rm g~cm^{-3}$\fi}
\def\gcmtwo{\ifmmode \rm g~cm^{-2} \else $\rm g~cm^{-2}$\fi}
   
\def\erg{\ifmmode {\rm erg} \else $\rm erg$ \fi}
\def\ergps{\ifmmode {\rm erg~s^{-1}} \else $\rm erg~s^{-1}$ \fi}
\def\ergcms{\ifmmode \rm erg~cm^{-2}~s^{-1} \else $\rm erg~cm^{-2}~s^{-1}$ \fi}
\def\ergcmshz{\ifmmode \rm erg~s^{-1}~cm^{-2}~Hz^{-1} \else $\rm
erg~cm^{-2}~s^{-1}~Hz^{-1}$ \fi}
\def\ergcmsa{\ifmmode \rm erg~cm^{-2}~s^{-1}~\AA^{-1} \else $\rm
erg~cm^{-2}~s^{-1}~\AA^{-1}$ \fi}
\def\ergshz{\ifmmode \rm erg s^{-1} Hz^{-1} \else
   $\rm erg s^{-1} Hz^{-1}$ \fi}

\def\lam{\ifmmode {\lambda} \else {$\lambda$} \fi}
\def\llam{\ifmmode {L_\lambda} \else  $L_\lambda$ \fi}
\def\lamLlam{\ifmmode \lambda L_{\lambda}(5100) \else {$\lambda L_{\lambda}(5100)$} \fi}
\def\nuLnu{\ifmmode \nu L_{\nu}(5100) \else {$\nu L_{\nu}(5100)$} \fi}
\def\lognuLnu{\ifmmode {\mathrm{log}\,\nu L_{\nu}(5100)} \else $\mathrm{log}\,\nu L_{\nu}(5100)$\fi}
\def\ilam{\ifmmode {I_\lambda} \else  $I_\lambda$ \fi}
\def\flam{\ifmmode {F_\lambda} \else  $F_\lambda$ \fi}
\def\inu{\ifmmode {I_\nu} \else  $I_\nu$ \fi}
\def\fnu{\ifmmode {F_\nu} \else  $F_\nu$ \fi}
\def\bnu{\ifmmode {B_\nu} \else  $B_\nu$ \fi}

\def\rblr{\ifmmode {R_\mathrm{BLR}} \else  $R_\mathrm{BLR}$\fi}

\def\logrm{\ifmmode {\mathrm{log}} \else ${\mathrm{log}}$\fi}
\def\msigma{\ifmmode M_{\sigma} \else $M_{\sigma}$\fi}
\def\mbulge{\ifmmode M_{\mathrm{bulge}} \else $M_{\mathrm{bulge}}$\fi}
\def\mgal{\ifmmode M_{\mathrm{gal}} \else $M_{\mathrm{gal}}$\fi}
\def\lgal{\ifmmode L_{\mathrm{gal}} \else $L_{\mathrm{gal}}$\fi}
\def\lbulge{\ifmmode L_{\mathrm{bulge}} \else $L_{\mathrm{bulge}}$\fi}
\def\mgalstar{\ifmmode M^*_{\mathrm{gal}} \else $M^*_{\mathrm{gal}}$\fi}

\def\mbhsigstar{\ifmmode M_{\mathrm{BH}} - \sigma_* \else $M_{\mathrm{BH}} - \sigma_*$ \fi}
\def\deltalogmbh{\ifmmode \Delta~{\mathrm{log}}~M_{\mathrm{BH}} \else $\Delta$~log~$M_{\mathrm{BH}}$\fi}

\def\sigstar{\ifmmode \sigma_* \else $\sigma_*$\fi}
\def\sigthree{\ifmmode \sigma_{\mathrm{[O~III]}} \else $\sigma_{\mathrm{[O~III]}}$\fi}
\def\sigtwo{\ifmmode \sigma_{\mathrm{[O~II]}} \else $\sigma_{\mathrm{[O~II]}}$\fi}
\def\signl{\ifmmode \sigma_{\mathrm{NL}} \else $\sigma_{\mathrm{NL}}$\fi}
\def\wthree{\ifmmode {\rm FWHM({[O~III]})} \else $FWHM({[O~III]})$ \fi}
\def\wtwo{\ifmmode {\rm FWHM({[O~II]})} \else $FWHM({[O~II]})$ \fi}
\def\mthree{\ifmmode M_{\mathrm [O~III]} \else $M_{\mathrm [O~III]}$ \fi}
\def\mtwo{\ifmmode M_{\mathrm [O II]} \else $M_{\mathrm [O II]}$ \fi}

\def\lbreak{\ifmmode L_{\mathrm{break}} \else $L_{\mathrm{break}}$\fi}
\def\lcut{\ifmmode L_{\mathrm{cut}} \else $L_{\mathrm{cut}}$\fi}

\slugcomment{Submitted to The Astrophysical Journal}

\shorttitle{Accretion Disk Temperatures of QSOs}
\shortauthors{Bonning et al.}

\begin{document}

\title{Accretion Disk Temperatures of QSOs:  Constraints from the Emission Lines}

\author{E.~W.~Bonning\altaffilmark{1,2}, G.~A. Shields\altaffilmark{3}, A.~C. Stevens\altaffilmark{3}, \& S.~Salviander\altaffilmark{3,4}}

\altaffiltext{1}{Quest University Canada, 3200 University Boulevard,
Squamish, BC, V8B 0N8, Canada; erin.bonning@questu.ca}

\altaffiltext{2}{Yale Center for Astronomy and Astrophysics, Yale University, P.O. Box 208121, New Haven, CT 06511, USA}

\altaffiltext{3}{Department of Astronomy, University of Texas, Austin,
TX 78712; shields@astro.as.utexas.edu, triples@astro.as.utexas.edu, alyx.stevens@mail.utexas.edu} 

\altaffiltext{4}{Department of Physics, Southwestern University, Georgetown, TX 78626}

\bibliographystyle{apj}

\begin{abstract}

We compare QSO emission-line spectra to predictions based on
theoretical ionizing continua of accretion disks. The observed line
intensities do not show the expected trend of higher ionization with
higher accretion disk temperature as derived from the black hole mass
and accretion rate. This suggests that, at least for accretion rates
close to the Eddington limit, the inner disk does not reach
temperatures as high as expected from standard disk theory.  Modified
radial temperature profiles, taking account of winds or advection in
the inner disk, achieve better agreement with observation. This
conclusion agrees with an earlier study of QSO continuum colors as a
function of disk temperature. The emission lines of radio-detected and
radio-undetected sources show different trends as a function of disk
temperature. 

\end{abstract}


\section{Introduction}

The idea of accretion disks around supermassive black holes as the power source of active galactic
nuclei (AGN) is widely accepted.   Physical plausibility, the lack of successful alternatives, and the frequent presence of bipolar jets, support
this picture.   Qualitatively, the blue-ultraviolet ``Big Blue Bump'' (BBB) 
in the spectral energy distribution of QSOs resembles the expected thermal emission from the disk atmosphere \citep{shields78, malkan83}.    
Models achieve some success in fitting the spectral energy distribution (SED) of
individual QSOs \citep[e.g.,][]{hubeny01}.  However, the simple disk model has known shortcomings in
accounting for the soft X-ray continuum, optical polarization,
variability \citep[see review by][]{koratkar99}.

In an effort to test the correspondence between disk theory and observation, 
\citet[hereinafter B07]{bonning07} examined the optical and ultraviolet colors 
of QSOs in the  Sloan Digital Sky Survey \footnote{The SDSS website is
http://www.sdss.org.} as a function of accretion disk 
temperature.   In simplest terms, do QSOs  with hotter disks show bluer colors?  They found poor
agreement between observed color trends and those predicted  by standard disk models.
For luminosities approaching the Eddington limit, 
the observed colors actually become redder with increasing disk temperature, 
in qualitative disagreement with predictions.  B07
suggested that the discrepancy might result from a modification of the inner disk within the radius where
radiation pressure gives a large fractional disk thickness.   Modified disk models with no emission inside
the critical radius gave improved agreement with observation.

Here we extend the results of B07 to include the QSO emission lines as diagnostics of the
higher frequency portion of the disk continuum.  The emission lines are governed by the intensity of the continuum at ionizing frequencies.
The emitted flux at these frequencies is more sensitive to the
accretion disk temperature and potentially offer a stronger test of the trend of the spectral energy distribution (SED) of QSOs with disk temperature.   
We again use the large dataset of quasar spectra available from the SDSS.  
We compare the observed quasar spectra to model predictions computed with the code AGNSPEC (Hubeny et al. 2000)
coupled with the photoionization code Cloudy \citep{ferland98}.   With a large range
of inferred disk temperatures, one might expect strong
trends in emission-line ratios and equivalent widths as a function of disk temperature.  
We find that these expectations are not borne out in observed QSO spectra. 

We refer the reader to B07 for further background and references to
earlier work.  We assume a concordance cosmology with 
$\Omega_{\Lambda}=0.7$, $\Omega_{m}=0.3$, and H$_{0}=70$ \kms Mpc$^{-1}$.

\section{Modeling and Data Sampling}
\label{sec:work}
\subsection{Accretion Disk Properties}
\label{sec:thorne}

B07 describe the physical basis of estimating the disk effective temperature.
 In standard thin-disk theory \citep{shakura73}, accreting matter gradually spirals inwards as viscous stresses 
 transport angular  momentum outward.  These stresses lead to local dissipation of energy that flows vertically to the surface of
the disk and is radiated by the photosphere.  The resulting effective temperature, given by
$F = \sigma\,\teff^4$, shows a radial dependence 
$\teff \propto r^{-3/4}$ at radii substantially larger than the inner boundary of the disk.
 As the inner boundary is approached, the effective temperature reaches a maximum
\tmax\ and then drops to zero at the inner boundary if the inner boundary condition is one of zero torque.  For
a disk radiating locally as a black body, the disk spectrum is fixed by \tmax, for a given value of black hole spin.
For realistic opacities, there are significant spectral features such as Lyman edges of hydrogen and helium,
either in emission or absorption \citep{hubeny01} that can cause disks with the same value of \tmax\ to
differ in their detailed continuum spectra.  However, \tmax\ still remains a useful sequencing parameter.

The key parameters for a QSO accretion disk are the black hole mass and spin,  the accretion rate, and the observer viewing angle.
An important reference luminosity is the Eddington limit, 
\begin{equation}
 L_\mathrm{Ed} = \frac{4\pi cG\mbh}{\kappa_{e}}=(10^{46.10}\, \ergps)M_8,
\end{equation}
where $\kappa_{e}$ is the electron scattering opacity per unit mass
and M$_8 = \mbh / (10^8~\mathrm M_{\odot})$.
The total luminosity produced is 
\begin{equation}
 L_{\rm bol} = \epsilon \dot{M} c^{2} = (10^{45.76}\, \ergps )\epsilon_{-1}
\mdoto.
\end{equation}
Here $\dot{M}_{\mathrm 0}$ is the accretion rate in \msunyr, and
$\epsilon$ = 10$\epsilon_{-1}$ increases from 0.057 for a non-rotating hole
to 0.31 for a rapidly rotating Kerr hole with angular momentum parameter $a_{\ast} = 0.998$  \citep{novikov73, thorne74}.   
For $a_{\ast}$ = 0.998
\begin{equation}
 \tmax=(10^{5.56}~{\rm K})M_{8}^{-1/4}(\lbol/\led)^{1/4},
\label{eq:tmax1}
\end{equation}
or alternatively,
  $\tmax=(10^{5.54}~{\rm K})M_{8}^{-1/2}\, L_{46}^{1/4}$, where
  $L_{46} \equiv \lbol/(10^{46}~\ergps)$. 
For a Schwarzschild hole, $T_\mathrm {max}$ is cooler by a factor of
$10^{0.46}$ for a given \mbh\ and \lbol\ \citep[e.g.,][]{shields89}.
We also define a reference accretion rate 
$\mdoted \equiv \led/c^2 =  (10^{-0.66}~\msunyr)M_8 $ and a dimensionless accretion rate
$\littlemdot \equiv \mdot/\mdoted$.   Note our definition of \mdoted\ in terms of $\epsilon = 1$,
so that a value $\littlemdot = 3.1$ gives \led\ for $a_{\ast} = 0.998$.

\subsection{Deriving \mbh\ and \tmax }
\label{sec:tmax}

Derivation of \mbh\ from the width of the \hbeta\ or \mgii\ broad
emission lines has become an accepted approximation in recent years.  The FWHM
of the broad lines is taken to give the circular velocity of the
broad-line emitting material (with some geometrical correction
factor).  The radius of the broad-line region (BLR), derived from echo
mapping studies, increases as a function of the continuum luminosity,
$R \propto L^{\Gamma}$.   \citet{bentz09} find $\Gamma = 0.52\pm0.07$, 
consistent with the value 0.5 suggested by photoionization physics.  
Here we use \citep{shields03}
\begin{equation}
\label{eq:mbh}
\mbh = (10^{7.69}~\msun)v_{3000}^2 L_{44}^{0.5},
\end{equation}
where $v_{3000}~\equiv~\rm{FWHM}$/3000~\kmps,  and
$ L_{44}~\equiv~\lamLlam/(10^{44}~\ergps)$. 

The derived value of \tmax\ depends almost entirely on the broad
line FWHM for a given object.  The bolometric luminosity can be
estimated as $\lbol = f_L \times \lamLlam$, following \citet{kaspi00}). Using  Eq.~\ref{eq:mbh} and this expression for \lbol,
Eq.~\ref{eq:tmax1} becomes
\begin{equation}
\label{eq:tmax2}
\tmax = (10^{5.43}~\mathrm{K})~v_{3000}^{-1}~L_{44}^{-(\Gamma -
0.5)/2}~(f_L/9)^{-1/4},
\end{equation}
We use a bolometric correction factor $f_L=9$ \citep{kaspi00, richards06}.

\subsection{SDSS Spectra and \tmax\ Bins }
\label{sec:bins}

We use the SDSS DR5 spectroscopic data base for ``quasars'' as our primary observational material.  We selected AGN using an automated spectrum fitting program and imposing quality cuts as described by \citet{salviander07}.   These fits give values for quantities such as the flux and width of the broad \hbeta\ line and the continuum flux at various rest wavelengths including $\lambda5100$.  The adopted redshift range is $z = 0.1$ to 0.8 so as to keep the [\oiii] nebular lines out of the telluric water vapor region.  Approximately one-half of the objects fall above the QSO/Seyfert galaxy luminosity boundary at $\lognuLnu = 44.44$ in units of \ergps.    Following B07, we derive \mbh\ and \tmax\ for the individual quasars as described above, and then group them into bins with \logtmax\ ranging from 5.0 to 5.7 defined by windows of 4.95 to 5.05, etc.  Table~1 gives the average values of \logmbh\  and \lognuLnu\ for these bins.  Also given is the accretion rate and resulting \logtmax\ on the assumption $f_L = 9$ and $a_\ast = 0.998$, based on the bin averages for \logmbh\ and \lognuLnu.  (An alternative approach to the accretion rate is discussed below.)     The values of \logmbh\ systematically decrease from 8.27 to 7.20 as \logtmax\ increases from 5.0 to 5.7.  The mean luminosity and accretion rate differs only slightly  among the bins.  As a result, the Eddington ratio increases with \logtmax\ from $\loglbyled = -1.5$ to -0.3 across the bins.

 \begin{deluxetable}{lcccccc}
 \label{tab:binprop}
 \tablewidth{0pt}
 \tablecaption{Properties \label{t:tab1}}
 \tablehead{
 \colhead{\tmax } &
 \colhead{\mbh} &
 \colhead{\nuLnu} &
 \colhead{$\mdot$} &
 \colhead{$L/\led$} &
 \colhead{$\mdot_{5100}$} &
 \colhead{$(L/\led)_{5100}$} \\
 \colhead{} &
 \colhead{ } &
 \colhead{ } &
 \colhead{ } &
 \colhead{}  
}
 \startdata  
5.0	 &	8.72		&	44.39		&	-0.95		&	 -1.47	&	-0.77 	&	-1.34\\	
5.1     &	8.51		&	44.36		&	-0.98		&	 -1.30	&	-0.63 	&	-0.99\\	
5.2	 &	8.31		&	44.33		&	-0.97		&	 -1.13	&	-0.49		&	-0.65\\
5.3	&	8.12		&	44.34		&	-0.95		&	 -0.93	&	-0.29 	&	-0.26\\
5.4	&	7.90		&	44.30		&	-1.00		&	 -0.75	&	-0.12 	&	 0.13\\
5.5	&	7.66		&	44.22		&	-1.06		&	 -0.59	&	-0.03 	&	 0.46\\
5.6	&	7.43		&	44.13		&	-1.13		&	 -0.44	&	 0.06 	&	 0.78\\
5.7	&	7.20		&	44.04		&	-1.19		&	 -0.30	&	 0.15 	&	 1.10\\	
 \enddata 
 \tablecomments{Mean properties of the SDSS quasars in the various \tmax\ bins.  
 All quantities are log$_{10}$, with \mbh\ in solar masses and \mdot\ in solar masses per year.  
Subscript ``5100'' refers to AGNSPEC models with accretion rate adjusted to match
observed continuum at 5100~\AA.  See text for discussion.}
 \end{deluxetable}

\begin{figure*}[]
\begin{center}
\includegraphics[scale=0.8]{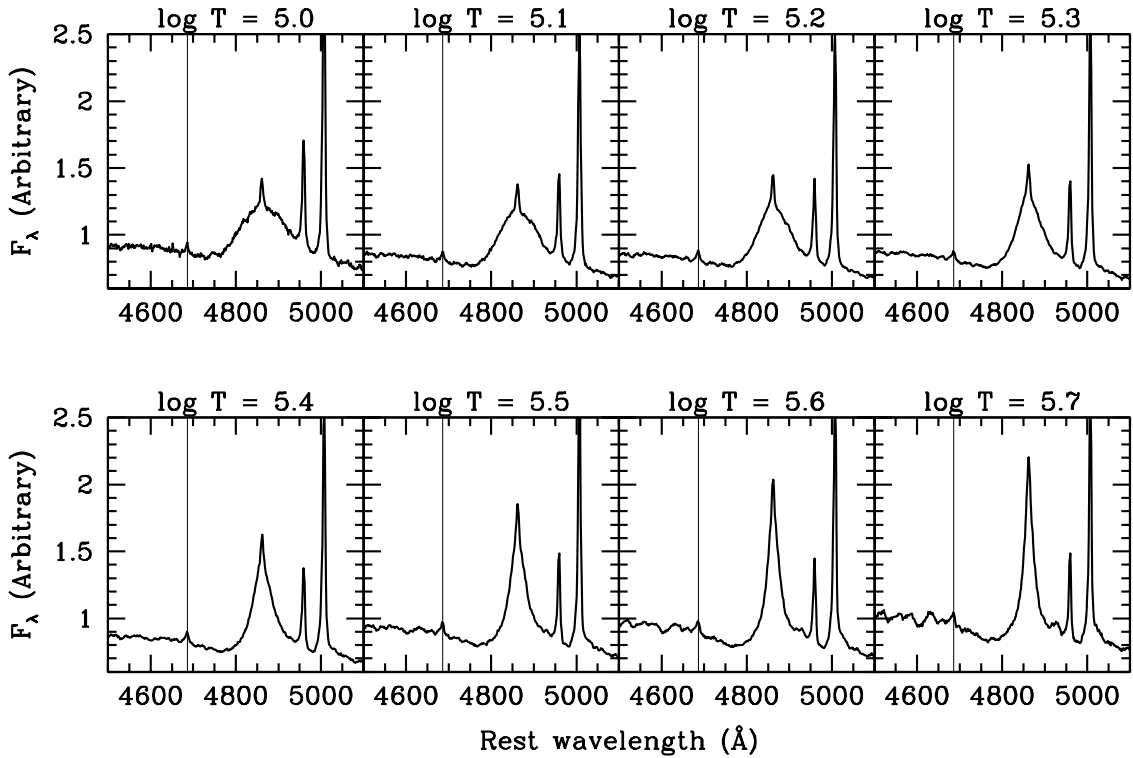}
\caption{Composite spectra of SDSS quasars binned by \tmax.  The thin vertical line represents the wavelength of the \heii~$\lambda4686$ line.  See text for discussion.}\label{fig:spectra}
\end{center}
\end{figure*}

\begin{figure}[]
\begin{center}
\includegraphics[scale=0.33]{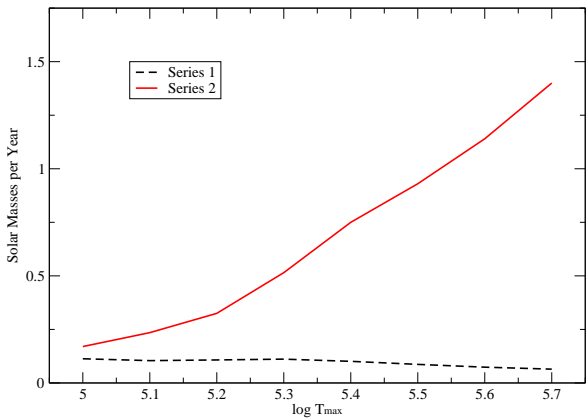}
\caption{Accretion rate used in the AGNPSEC models of the accretion disk. Lower curve (black dashed) is the accretion rate based on an assumed bolometric correction $f_L = 9$. Upper curve (red solid) shows the accretion rate required to give the mean observed $\lambda5100$ continuum luminosity of the objects in the \tmax\ bins. For the higher values of \tmax, the bolometric correction is large and the accretion rate and resulting \tmax\ is higher than for $f_L = 9$. This aggravates the discrepancy between predicted and observed emission-line properties.}\label{fig:mdot}
\end{center}
\end{figure}

For each of the bins, we form a composite spectrum giving equal weight to all objects based on the frequency-integrated flux at earth over the spectrum.   These composite spectra were measured to give values of several observational quantities (e.g., the [\nev] emission-line intensity) used in the analysis below.  Line measurements were carried out in IRAF\footnote{IRAF is distributed by the National Optical Astronomy Observatories, which are operated by the Association of Universities for Research in Astronomy, Inc., under cooperative agreement with the National Science Foundation.}. Figure \ref{fig:spectra} gives the composite spectra and Figure \ref{fig:mdot} shows the accretion rate for the composite spectra as discussed in the preceding paragraph.

%\begin{figure*}[]
%\begin{center}
%\plotone{fig1.eps} 
%\figcaption[fig1]{ Composite spectra of SDSS quasars binned by \tmax.  The thin vertical line represents the wavelength of the \heii~$\lambda4686$ line.  See text for discussion.
%\label{fig:spectra} }
%\end{center}
%\end{figure*}

%\begin{figure}[]
%\begin{center}
%\plotone{fig2.eps} 
%\figcaption[fig1]{ 
%Accretion rate used in the AGNPSEC models of the accretion disk. Lower curve (black dashed) is the accretion rate based on an assumed bolometric correction $f_L = 9$. Upper curve (red solid) shows the accretion rate required to give the mean observed $\lambda5100$ continuum luminosity of the objects in the \tmax\ bins. For the higher values of \tmax, the bolometric correction is large and the accretion rate and resulting \tmax\ is higher than for $f_L = 9$. This aggravates the discrepancy between predicted and observed emission-line properties.  
%\label{fig:mdot} }
%\end{center}
%\end{figure}

\subsection{Accretion Disk Models}
\label{sec:bins}

Our goal is to compare basic trends in observed and predicted AGN properties as a function of \tmax.  To this end, we have computed disk models using the AGNSPEC program \citep{hubeny01} in the manner described in B07.  These models take as input parameters the black hole mass
\mbh\ and spin paramter $a_\ast$, accretion rate \mdot, viscosity parameter $\alpha$, and inclination $i$ of the disk axis to the line of sight.    For simplicity, we adopt the commonly assumed value $\alpha = 0.1$ and focus on rapidly rotating black holes with $a_\ast = 0.998$ \citep{thorne74}.  We take $\mathrm{cos}\,i = 0.6$ as representative of randomly oriented AGN subject to an exclusion of inclinations near edge-on because of the dusty torus postulated in the unified model  of AGN \citep{antonucci85}.  The AGNSPEC code computes the
vertical gravity and locally emitted flux $F(R)$ in the orbiting frame, and calculates for each radius the vertical structure and emergent spectrum with the aid of a pre-computed grid of models.  The spectrum observed at infinity is  summed over all emitting radii accounting for
all relativistic effects.  In order to avoid an excessive number of free parameters, we restrict our models to the case $a_\ast = 0.998$.  The spin of
the black holes in AGN has been much debated, with some authors arguing for moderate values based on likely merger histories or
fitting of disk models to the spectral energy distributions of individual AGN \citep[e.g.,][and references therein]{czerny11}.

\section{Problems with the Standard Disk Model}
\label{standard}

\begin{figure}[]
\begin{center}
\includegraphics[scale=0.33]{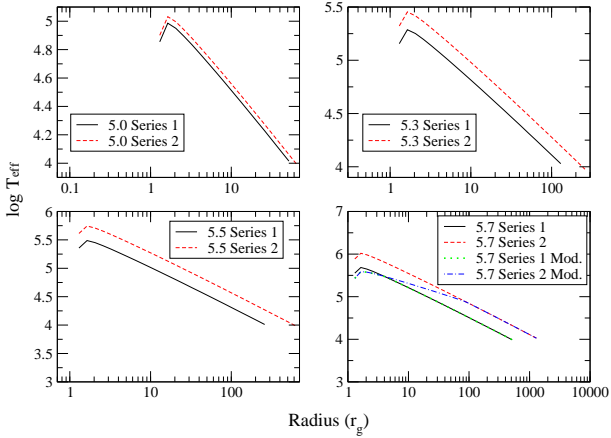}
\caption{Radial temperature profile for the standard disk model and the luminosity fitted model, including the alternative models with a modified temperature profile in the inner disk.  See text for discussion.}\label{fig:modified}
\end{center}
\end{figure}

Here we discuss tests of the success of the standard thin accretion disk model in explaining a variety of observed properties of AGN. 
We discuss two different approaches to determining the accretion rate of the disk, which together with \mbh\ determines \tmax.  First, we
follow B07 in assuming a universal bolometric correction $f_L = 9$ as described above.  Then we discuss an alternative approach (see Section 3.3) in which \mdot\ is adjusted in the disk models so as to reproduce the observed value of \nuLnu. Figure \ref{fig:mdot} shows the accretion rate derived for the composite spectra in these two different approaches. Figure \ref{fig:modified} shows the radial temperature profile for these models as well as a modified accretion disk model discussed in Section 4.

%\begin{figure}[]
%\begin{center}
%\plotone{fig3.eps} 
%\figcaption[fig1]{
%Radial temperature profile for the standard disk model and the luminosity fitted model, including the alternative models with a modified temperature profile in the inner disk.  See text for discussion. 
%\label{fig:modified} }
%\end{center}
%\end{figure}

\subsection{Emission-Line Intensities}

 As found above, typical values of \tmax\ for the accretion disk (for
the parameters of our sample)  are of order 200,000~K,  
 corresponding to a black body peak at $h\nu \approx 3kT \approx  1.5 h\nu_{\mathrm H}$, where
$h\nu_{\mathrm H} = I_{\mathrm H}$ is the hydrogen ionization potential.   Thus, the output of the disk in hydrogen
ionizing photons, $Q(\mathrm{H}^0)$, will be sensitive to the value of \tmax, and the sensitivity will be greater for ions with higher ionization potentials.
The amount of ionized gas and the resulting line luminosities, as well as the relative intensities of lines
from different levels of ionization, should depend strongly on \tmax.

\subsubsection{\hbeta\ Equivalent Width}
\label{sec:hbew}

A basic application of this concept is the equivalent width of the broad \hbeta\ line.  For a photoionized nebula, 
the  luminosity of the \hbeta\ line is given by radiative recombination theory \citep{osterbrock06} as
\begin{equation}
\label{eq:ewhb}
L(\hbeta) = (\Omega/4\pi) (\epsilon_{\hbeta}/\alpha_{\mathrm B}) Q(\hzero).
\end{equation}

\begin{figure}[]
\begin{center}
\includegraphics[scale=0.33]{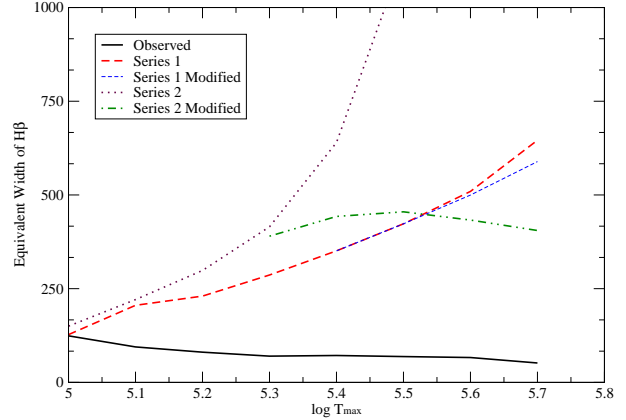}
\caption{Predicted and observed equivalent width of broad \hbeta\ line as a function of \tmax. The SDSS composites show a systematic decrease in EW with increasing  \tmax, in contrast to the strong increase predicted by the standard disk model. The modified disk model shows a leveling off of the EW for higher \tmax\ although at an excessively large value; the discrepancy in slope remains at lower \tmax}\label{fig:ewhbeta}
\end{center}
\end{figure}

Here, $\Omega/4\pi $ is the covering fraction for the BLR, $\epsilon_{\hbeta}$ is the recombination-line emission
coefficient, $\alpha_B$ is the radiative recombination coefficient for hydrogen to quantum levels $\mathrm{n} = 2$ and higher,
and $Q(\hzero)$ is the ionizing photon luminosity (photons per second).
This does not allow for collisional excitation or other processes that can affect \hbeta\ in the BLR.
The \hbeta\ equivalent width is then given by $\mathrm{ EW} = L(\hbeta)/\llam(4861 \, \AA)$.  As
\tmax\ increases, we expect a strong increase in  $\mathrm{ EW(\hbeta)}$, because the ionizing frequencies are
more sensitive to disk temperature than the optical continuum emission.  Figure \ref{fig:ewhbeta} shows the predicted EW 
for the set of AGNSPEC models described above, on the assumption of a covering fraction of 0.5 independent of
\tmax.  The expected increase of EW with \tmax\ is evident.  (The covering fraction of 0.5 was chosen to give agreement
with the observed EW of \hbeta\ at log~\tmax = 5.0 and is reasonable
in the context of the unified model of AGN.)

%\begin{figure}[]
%\begin{center}
%\plotone{fig4.eps} 
%\figcaption[fig1]{ Predicted and observed equivalent width of broad \hbeta\ line as a function of \tmax.   The SDSS composites show a systematic decrease in EW with increasing  \tmax, in contrast to the strong increase predicted by the standard disk model.   The modified disk model shows a leveling off of the EW for higher \tmax\ although at an excessively large value; the discrepancy in slope remains at lower \tmax.
%\label{fig:ewhbeta}}
%\end{center}
%\end{figure}

Figure \ref{fig:ewhbeta} also shows the observed trend of the \hbeta\
EW as a function of \tmax\ for our SDSS composites.  The observed
values show a progressive {\em decrease} from from 120~\AA\ to 50~\AA\
as \logtmax\ increases from 5.0 to 5.7.  This trend is in qualitative
disagreement with the model prediction.  While it is possible that the
covering fraction changes systematically with \tmax, an
order-of-magnitude decrease in $\Omega/4\pi$ would be required from
$\logtmax = 5.0$ to 5.7.  A search for independent signatures of such
a trend may be worthwhile.  However, it seems likely that such an
extreme trend in covering fraction would have been been noticed in
prior investigations.  Therefore, the discrepancy shown in Figure
\ref{fig:ewhbeta} is an indication that something is wrong with the
standard disk model as regards the ionizing photon output.  This
discrepancy resembles that found in B07 in which the continuum colors
do not show the expected trend toward bluer colors at higher \tmax.
The systematic decrease in \hbeta\ equivalent width with increasing
\tmax\ gives a new context to the long known correlation between the
EW and FWHM of the broad \hbeta\ emission line \citep{boroson92},
since the derived \tmax\ depends mainly on the \hbeta\ broad line
width (see eq. 5).

\subsubsection{Ionization Ratios}
\label{sec:ratios}

Typical AGN spectra show emission lines from ions with a large range of ionization potentials.  
In the case of ionized nebulae,
a useful diagnostic of the ionizing continuum is the ratio of the lines of He~II and H~I.  In a normal nebular ionization structure,
\heii\ occupies an inner volume of the Str\"omgren sphere  (or a surface layer of a slab) whose fraction of the
total H$^+$ volume is proportional to the ionizing photon luminosity ratio, $Q(\heplus)/Q(\hzero)$.  Here 
$Q(\heplus)$ is the ionizing photon luminosity above 4 Rydbergs frequency.  The recombination line ratio
$I(\heii\,\lambda4686)/I(\hbeta)$ is in turn proportional to the nebular average of $N(\heplustwo)/N(\hplus)$, giving
\begin{equation}
\label{eq:i4686}
I(\lambda4686)/I(\hbeta) = 1.02\,  Q(\heplus)/Q(\hzero).
\end{equation}
This assumes that the nebula is optically thick to the Lyman continuum.

%\begin{figure}[]
%\begin{center}
%\plotone{fig5.eps}
%\figcaption[fig1]{ 
%Predicted line intensity ratio of  \heii/\hbeta\  for accretion disk model compared with observed ratio for the narrow lines as a function of \tmax.  Note strong underprediction of \heii\ for the lowest temperatures and the strong trend in the model predictions in contrast to the observed intensities.  The \heii\ narrow line intensity is difficult to separate from the broad line profile for higher \tmax.  See text for discussion.
%\label{fig:qhe}}
%\end{center}
%\end{figure}

\begin{figure}[]
\begin{center}
\includegraphics[scale=0.33]{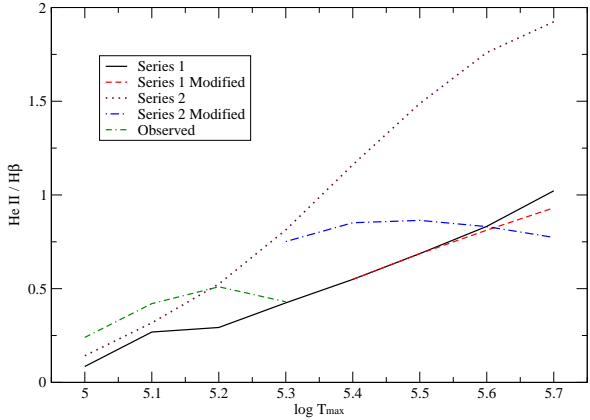}
\caption{Predicted line intensity ratio of  \heii/\hbeta\  for accretion disk model compared with observed ratio for the narrow lines as a function of \tmax.  Note strong underprediction of \heii\ for the lowest temperatures and the strong trend in the model predictions in contrast to the observed intensities.  The \heii\ narrow line intensity is difficult to separate from the broad line profile for higher \tmax.  See text for discussion.}\label{fig:qhe}
\end{center}
\end{figure}

Figure \ref{fig:qhe} shows that $Q(\heplus)/Q(\hplus)$ does indeed increase strongly with increasing \tmax\ in the AGNSPEC models. 
Figure \ref{fig:spectra}  shows the composite spectra for the various bins in \logtmax.   The broad \heii\ line is blended with
\feii\ and difficult to measure.  The \heii\ line in the narrow line spectrum is well defined for our cooler composites, in which
the broad line widths are large.  However, it is difficult to separate the narrow from the broad components of
$\lambda4686$ for the hotter composites, which have relatively narrow broad lines.   Nevertheless, two points can be be made. A subjective inspection of the figure indicates little trend in the narrow \heii\ strength as a function of \logtmax.  Also, the observed values of
narrow \heii/\hbeta\ for the lower \tmax\ bins are larger than would be expected for the predicted disk continuum, absent an additional hard component such as a power-law; and there is little trend with increasing \tmax.

\begin{figure}[]
\begin{center}
\includegraphics[scale=0.33]{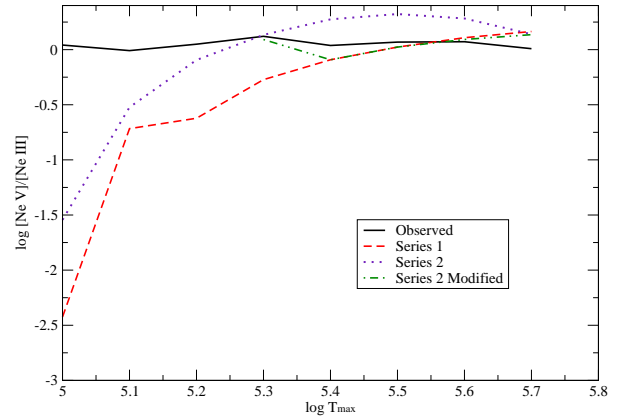}
\caption{Predicted line intensity ratio of  \nev/\neiii\  for accretion disk model compared with observed ratio as a function of \tmax.  Note strong trend in the model prediction that is not reflected in the observed intensities.   See text for discussion.}\label{fig:neon}
\end{center}
\end{figure}

For an observationally more feasible test, we consider the
narrow line ratio of [\nev]~$\lambda3426$ to [\neiii]~$\lambda3869$.   The \neplusfour\ ion requires even higher ionizing photon energies than
\heplustwo\ and thus the [\nev]/[\neiii] ratio should be even more sensitive to \tmax\ than \heii/\hbeta.
However, [\nev]/[\neiii] shows no significant trend with \tmax\ (see Figure \ref{fig:neon}). Thus, this observed line
ratio, involving emission lines from the NLR rather than the BLR, also fails to support the expectation 
of a harder spectrum in the ionizing ultraviolet for higher \tmax.

%\begin{figure}[]
%\begin{center}
%\plotone{fig6.eps} 
%\figcaption[fig1]{ 
%Predicted line intensity ratio of  \nev/\neiii\  for accretion disk model compared with observed ratio as a function of \tmax.  Note strong trend in the model prediction that is not reflected in the observed intensities.   See text for discussion.
%\label{fig:neon}}
%\end{center}
%\end{figure}
 
\subsection{Soft X-rays}
\label{sec:xray}

\begin{figure}[]
\begin{center}
\includegraphics[scale=0.33]{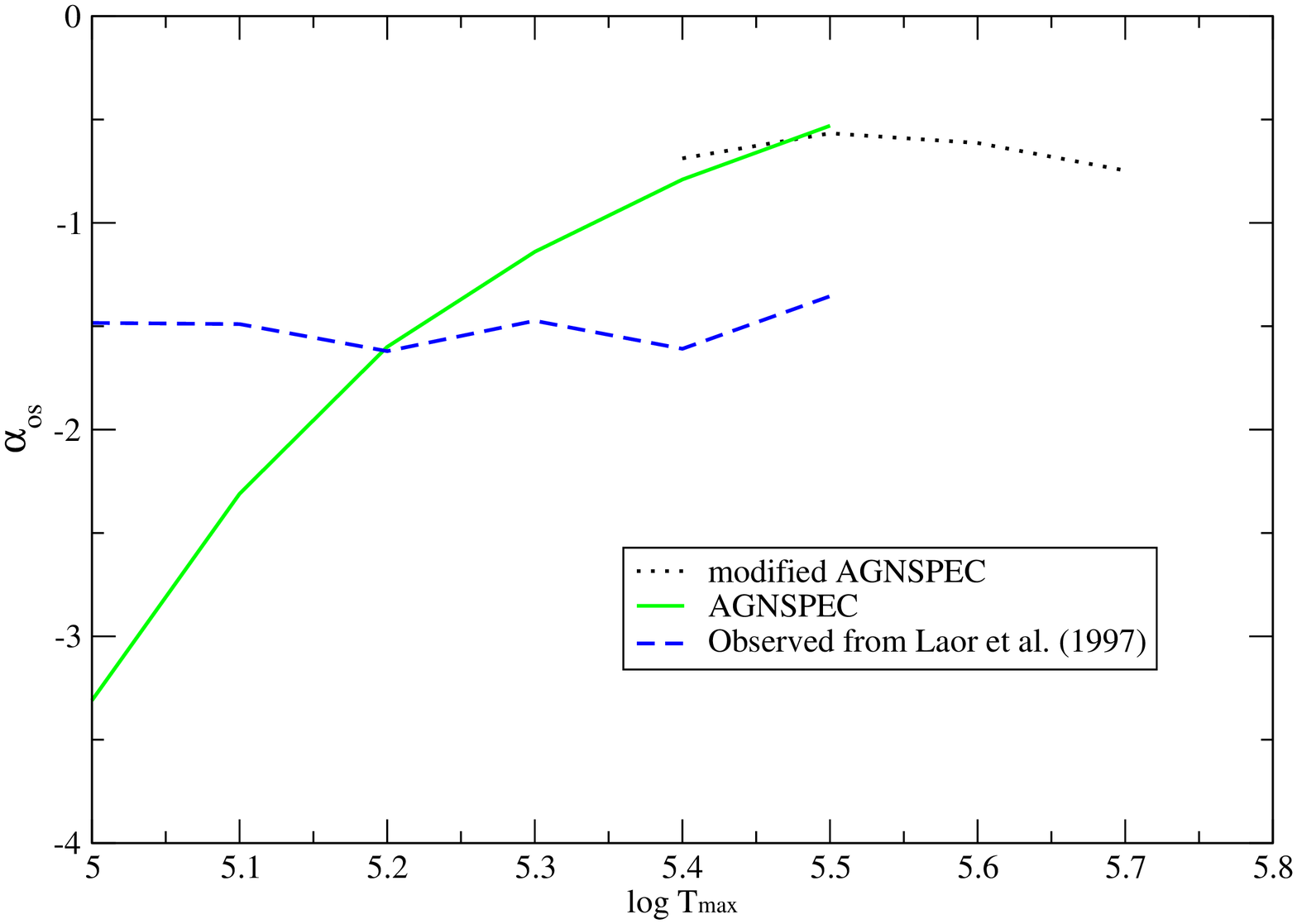}
\caption{Predicted optical to soft X-ray spectral index \alphaos\ from AGNSPEC models compared with observed values derived from \citet{laor97}. Note absence of trend with \tmax\ for the observed values.  See text for discussion.}\label{fig:alphaos}
\end{center}
\end{figure}

AGN spectra characteristically include emission in the hard and soft X-ray bands.  In general, this emission involves frequencies too high to be the result of ordinary thermal continuum emission from a standard accretion disk.  \citet{laor97} and others have noted the inconsistency of observed
soft X-ray fluxes with the simple thin-disk model.  A Comptonizing corona, producing a power-law spectrum extending to high photon energies, is sometimes invoked \citep{czerny87}.  However, for the higher values of \tmax\ that we consider here, substantial soft X-ray emission from the inner disk is predicted, and there is the possibility of overproducing soft X-rays. 
We have computed \tmax\ in the manner described above, for the AGN in the sample described by \citet{laor97}.  Laor gives X-ray fluxes and values of the optical (3000~\AA) to soft X-ray (0.3~keV) spectral index defined by
$L_{\mathrm s}/L_{\mathrm o} = (\nu_s/\nu_o)^\alphaos$.  
Figure \ref{fig:alphaos} shows the predicted value of  $\alphaos$
as a function of \tmax\ for a simple set of AGNSPEC models with $\mbh = 10^8~\msun$.  
The increase in \alphaos\ with increasing \tmax\ is evident.  In contrast, the observed values of \alphaos\
in the Laor sample, grouped into our same bins in \tmax, show no significant trend with \tmax.  In particular, for
$\logtmax > 5.2$, the models predict a greater value of $L_{\mathrm s}/L_{\mathrm o}$ than observed.  This is
significant, because one might invoke a Comptonized component to explain the soft X-rays for low
\tmax; but the hotter disks, in the standard model, over-predict the soft X-rays even before any
Comptonized component is added.  While the origin of the X-ray flux of AGN remains uncertain, it is remarkable that
observed values of \alphaos\ show no trend with \tmax.  Thus, we have in \alphaos\ yet another qualitative discrepancy
between observed AGN properties and the predictions of the simple disk model.

%\begin{figure}[]
%\begin{center}
%\plotone{fig7.eps} 
%\figcaption[fig1]{ 
%Predicted optical to soft X-ray spectral index \alphaos\ from AGNSPEC models compared with observed values derived from \citet{laor97}. Note absence of trend with \tmax\ for the observed values.  See text for discussion.
%\label{fig:alphaos} }
%\end{center}
%\end{figure}

The dependence of \alphaos\ on disk effective temperature has been previously discussed by other authors.  For example,  \citet{deroches09} in their Figure 6  show \alphaos\ versus $\mathrm{log} [(\lbol/\led)^{1/4} \mbh^{1/4}]$.   Aside from minor differences in the derivation of \lbol\ and \mbh, their temperature axis can be calibrated to our \logtmax\ scale by adding 7.56.  The SDSS objects in their plot span a similar range in \tmax\ to ours, and likewise show little systematic trend in \alphaos\ with \tmax.

\subsection{Models with variable bolometric correction}

As noted above and in B07, the use of a universal bolometric correction $f_L = 9$ is not consistent with the proposition that disk temperatures may differ considerably among AGN.  An alternative approach is to tune the value of \mdot\ in the AGNSPEC models so as to give the
observed value of  \lamLlam\  for a given \tmax\ bin.  We have computed a set of models in this fashion, still assuming
$a_\ast = 0.998$ and $\mu = 0.6$.  We find that for $\logtmax = 5.0$, the resulting model has a value  \lbol\ giving $f_L\approx 10$, so that the
model differs little from the $f_L = 9$ case.  However, for the higher \tmax\ bins, the value of $f_L$ implied by the AGNSPEC fits increases progressively to a value of $f_L \approx 50$ for $\logtmax = 5.7$.  (Here we use the terminology \tmax\  to refer to the original SDSS quasar bins, even though the
actual \tmax\ of the models is now higher.)  These models involve very large accretion rates which exceed the more conventional \mdot\ based on $f_L = 9$ by nearly an order of magnitude.  Such large accretion rates put increased strain on  available
accretion supplies.  They also give \lbol\ much larger than \led, so that the inner radii of such disks would likely involve serious modifications due to
radiation pressure and mass loss, as discussed below.   Nevertheless, we start our discussion of these models by revisiting the several observational tests above using the predicted disk spectra from AGNSPEC.  

Because $f_L$ increases strongly with \tmax\ in these models, the actual value of \tmax\ exceeds the nominal value defining the data bin by an increasing factor as the nominal \logtmax\ increases.   These models are useful, nevertheless, because we may think of the nominal \logtmax\ spectrum bins as a sequencing procedure that gives us a set of SDSS AGN involving a large range in \tmax, even if there are ambiguities as to how well the nominal value agrees with the actual disk temperaure.   Because the models adjusted to $L_{5100}$ have a greater range of \mdot\ and \logtmax\ than the nominal models,  they are expected to show even stronger trends in observational quantities sensitive to \tmax, such as
continuum color and high ionization emission-line strength.

 Figure \ref{fig:ewhbeta} shows the predicted equivalent width of \hbeta\ for the luminosity-fitted models.  The strong increase with the bin value of  \tmax\ is now even more pronounced and in greater contrast to the observed decrease.  Figures \ref{fig:qhe} and \ref{fig:neon} 
 show the predicted \heii\ and  [\nev] intensity.   The strong increase in [\nev] with
 \tmax\ is now in even greater contrast to the observed constancy of [\nev]/[\neiii].  Figure \ref{fig:alphaos} shows the extreme trend in \alphaos\ versus \tmax\ for
 the luminosity-fitted models.  The hotter cases greatly overpredict the observed soft X-ray fluxes.  

%xxx fig6
 
 Clearly, the luminosity-fitting approach only
 exacerbates the discrepancy between the theoretical predictions and observations.   However, this approach may have merit physically, since it avoids
 use of a universal bolometric correction.  Moreover, for the higher values of \tmax, the $f_L = 9$ models severely underpredict the luminosity
 at $\lambda5100$.   If the accretion rate is not increased, some restructuring of the AGN continuum source is needed to channel energy 
 more efficiently into the continuum at optical wavelengths.   In any case, the standard thin-disk model requires some kind of modification to
 explain the continuum and emission-line observations.

\section{A Modified Disk Model}

B07 found that observed continuum colors of SDSS QSOs did not become progressively bluer with increasing \tmax\ in the manner predicted by the AGNSPEC models.  We suggested that this resulted from higher Eddington ratios at higher \tmax, leading to departures from the standard
thin disk model.  This might involve thickening by radiation pressure, inefficient ``slim disk'' accretion, and mass loss in a radiation-driven wind (see also Deroches et al. 2009).  As an
estimate of the potential effect on the continuum colors, we computed AGNSPEC models in which the inner disk was truncated inside the
radius where radiation pressure gave a vertical scale height $H$ for the disk, relative to the radius, of $H/R = 0.5$.  This led to improved agreement with the
observed colors, in particular giving a non-monotonic trend of color with \tmax.

Here we explore a more physical model in which the disk effective temperature is modified inside the critical radius.  The basic idea,
following \citet{poutanen07} and \citet{begelman06}, is to take account of the luminosity produced at a given radius relative to
\led.  If the luminosity emitted by a wide annulus, \dlogLR, is well below \led, then the disk will not
be severely thickened by radiation pressure, and standard disk physics applies.    However, except near the inner boundary,
\dlogLR\ increases with decreasing $R$.  For sufficiently large values of $\mdot/\mbh$, a radius is reached where \dlogLR\ reaches \led.
Near this radius, serious modifications of the disk structure can be anticipated.   \citet{shakura73} called this the ``spherization
radius''  \rsp\ on the basis that radiation pressure would severely thicken the disk.    \citet{poutanen07} find that,
inside  \rsp, a combination of wind mass loss and advection occurs to a degree that keeps the local luminosity \dlogLR\ close to \led.  
This results in a break in the run of effective temperature with radius, so that inside \rsp\ one has $\teff \propto R^{-1/2}$ 
rather than the standard dependence $\teff \propto R^{-3/4}$.  This reduces the value of \tmax\ actually reached in the inner disk
as well as the luminosity radiated from the hottest parts of the disk.

The value of \rsp\ in units of the gravitational radius is approximately $\littlersp \equiv \rsp/\rgrav \approx \littlemdot.$
As \mdot\ increases, the effective temperature at \rsp\ actually decreases, following 
\begin{equation}
\tsp = (10^{5.75}{\rm~K})\littlemdot^{-1/2}M_8^{-1/4}.
\end{equation}
Here we have used the Newtonian expression for the local flux, omitting the normal factor of 3 increase in flux due to viscous transport of energy from smaller radii.  This modification was found by \citet{poutanen07} because of energy loss in the inner disk due to advection and mass loss.   For large values of \littlemdot, 
a luminosity $\sim\led$ is emitted in the vicinity of \rsp, and a similar luminosity is emitted per unit $\mathrm{ln}~R$ at smaller radii.  Thus \tsp\ substantially
controls the character of the emitted spectrum.   Hence, the fact that \tsp\  {\em decreases} with increasing \littlemdot\  once spherization comes into play
is suggestive of the reversal in the trend of colors with \tmax\ found in B07.  (Note that, in this context, \tmax\  is defined in terms of the accretion rate
entering the disk at large radii, and differs from the actual maximum temperature in the inner disk.)

The potential for an inverse relationship between \mdot\ and disk color temperature is strengthened if one considers the optical depth of the disk wind.
In the case of highly super-Eddington accretion, \citet{poutanen07} find that a large part of the inflowing material is expelled as a wind in the vicinity of \rsp.
In the simple approximation of a spherical wind with constant radial velocity equal to the circular orbital velocity at \rsp, the optical depth
of the wind, measured radially from \rsp, is $\tau_\mathrm{sp} \approx \littlemdot^{1/2}$.  The optical depth  in the wind, measured outward from some
radius $R$ that exceeds \rsp, is $\tau(>R) \approx \tau_\mathrm{sp} (R/\rsp)^{-1}$.  For the radius of the photosphere, where $\tau(>R) = 1$,
this gives $\littlerph \approx \littlemdot^{3/2}$.  The effective temperature at the photosphere is
\begin{equation}
\tph = (10^{5.75}{\rm~K})\littlemdot^{-3/4}m_8^{-1/4}.
\end{equation}
This raises the prospect that, for super-Eddington accretion rates, the maximum visible disk temperature will be  \tph\ 
rather than \tmax.  An estimate of the progression of the disk temperature as it actually
appears to the observer is then
\begin{equation}
\tapp = \mathrm{min}(\tmax, \tph).
\end{equation}
This assumes that there is enough absorption opacity in the wind to reprocess the emerging energy; otherwise, the escaping spectrum may
still be characterized by \tsp.

For our Series 1 models above, with an accretion rate based on $f_L = 9$, the value of  \loglbyled\ reaches -0.3 for 
$\logtmax = 5.7$.  Thus, unless opacity sources other than electron scattering reduce the effective Eddington limit, the effects of
radiation pressure may be modest.  However, for our Series 2 models with \mdot\ adjusted to fit the continuum at $\lambda5100$,
\loglbyled\ reaches +1.1 for \logtmax\ = 5.7.    For this series,  at $\logtmax=(5.0, 5.4, 5.7)$ we have $\mathrm{log}\, \littlemdot=(-0.79, +0.29, +1.65)$;
$\mathrm{log}\,  \mathrm{min}(\tmax, \tsp)=(5.05, 5.44, 5.13)$, and $\mathrm{log}\, \mathrm{min}(\tmax, \tph)=(5.05, 5.37, 4.75)$.
Thus, a reversal of the trend of the apparent temperature of the disk as a function of the nominal \tmax\ would be expected at 
$\logtmax \approx 5.4$.  The fact that this value agrees with the turn-around in the color trend found in B07 is fortuitous, given
the approximations in our discussion.  However, this simple calculation does illustrate the possibility of reverse trend of
color temperature with $\mdot/\mbh^2$ for super-Eddington rates.

In order to explore the impact of super-Eddington accretion on the disk spectrum and resulting emission-line intensities, we have computed modified models with AGNSPEC in which the standard \teff\ inside a break radius $R_b$ is multiplied by a factor $(R/R_b)^{1/4}$ (see above).  
For the value of the break radius, we took $R_b/r_g = x_b\,\,(L/\led)$.  We present here results for $x_b = 7$, which was motivated by a
preliminary analysis but which serves here simply as an illustrative example.  We used the same values of \mbh\ and \mdot\ as used in our Series 1 and Series 2 models above. The run of \teff\ with radius for the modified models is shown in Figure \ref{fig:modified}.  As expected, there is little effect for  Series 1 ($f_L = 9$); but substantial alterations are seen for Series 2. In Figure \ref{fig:ewhbeta}, the Series 2 prediction for the \hbeta\ equivalent width levels off above $\logtmax = 5.4$.   This reduces but does not eliminate the discrepancy with the observed trend.   Figure \ref{fig:qhe} shows a similar leveling of the predicted \heii\ intensity.  Figure \ref{fig:neon} shows a substantial reduction in the predicted \nev\ intensity for \logtmax\ around 5.4.  These models are not intended to achieve a detailed fit to the observed spectra, but they do illustrate the substantial effect on the ionizing continuum that results from a modified disk structure resulting from super-Eddington accretion.

\section{Radio vs. Non-radio sources}
\label{sec:radio}

Our focus has been on the entire SDSS quasar data set, which is strongly dominated by radio quiet objects.  However, radio loudness 
raises interesting questions regarding the black hole spin, which affects the inner disk boundary and thus the maximum temperature reached in the
disk.  Numerous authors have discussed the possibility that radio loud AGN have rapidly spinning black holes \citep[e.g.,][]{blandford90},
and on the other hand \citet{garofalo10} have proposed that radio loud objects have {\em retrograde} black hole spin with respect to the
disk angular momentum.
The innermost stable circular orbit (ISCO) decreases from $6 R_g$ for $a = 0$ to $1.22 R_g$ for $a_\ast = 0.998$.  Correspondingly,
the radius of maximum effective temperature moves inward from $9.5 R_g$ to $1.55 R_g$; and the value of \tmax\ increases
from $10^{4.79}\mdoto^{1/4} M_8^{-1/4}$ to $10^{5.43 }\mdoto^{1/4} M_8^{-1/4}$ \citep{shields89}.  
For retrograde spin, the ISCO moves still farther out.  These differences in
the inner disk temperature could have a significant effect on the ionizing spectrum of the disk, while having less effect on the
optical luminosity.   Thus, we might expect differences between radio loud and quiet AGN in the observational quantities sensitive to \tmax\ discussed above.   

%\begin{figure}[htb!]
%\begin{center}
%\plotone{fig8.eps} 
%\figcaption[fig1]{ 
%Comparison of the [\nev]/[\neiii]  ratio for radio detected and non-detected SDSS quasars as a function of \tmax.  Note significant difference for the higher values of \tmax.  See text for discussion.
%\label{fig:radio} }
%\end{center}
%\end{figure}

\begin{figure}[]
\begin{center}
\includegraphics[scale=0.33]{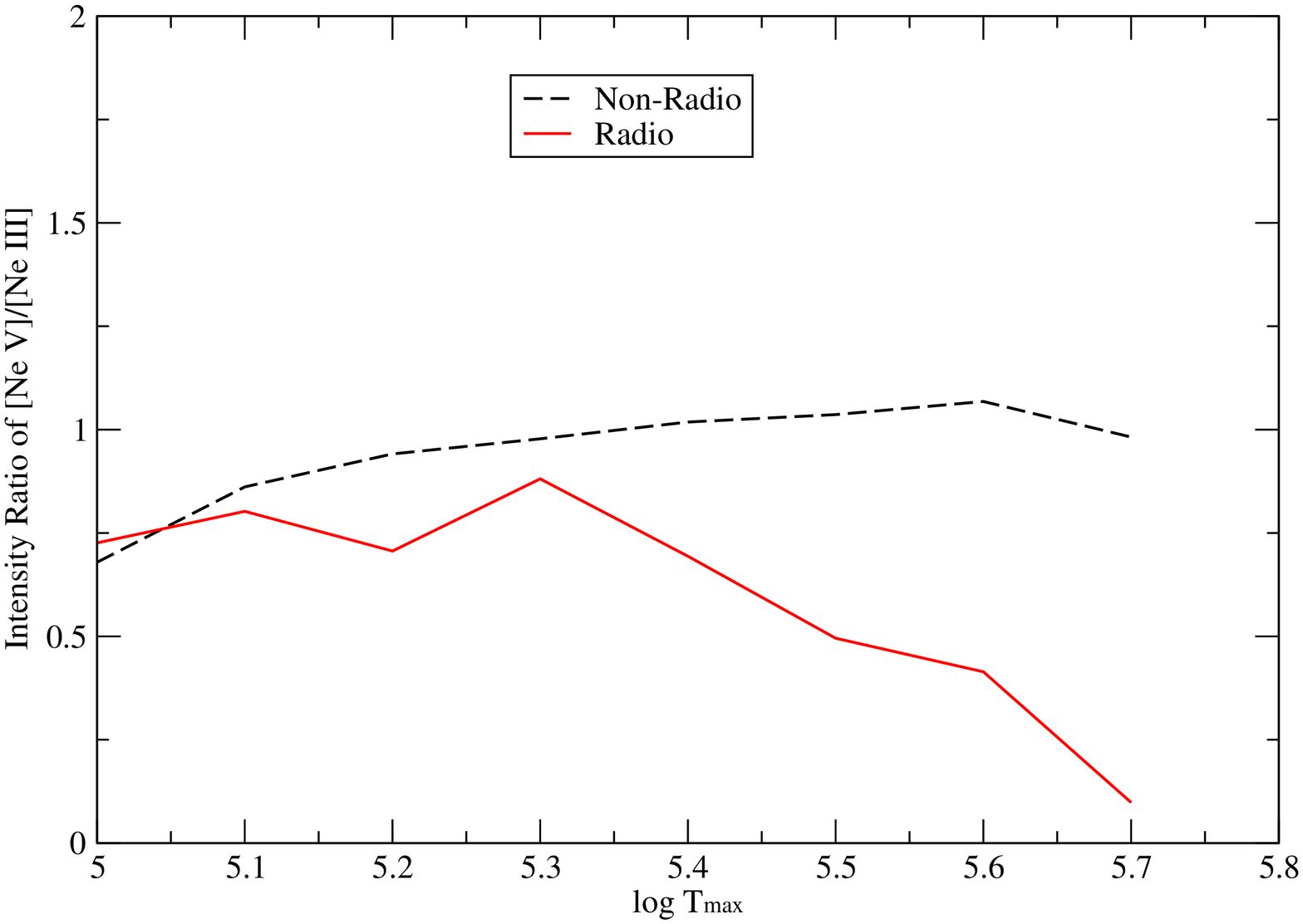}
\caption{Comparison of the [\nev]/[\neiii]  ratio for radio detected and non-detected SDSS quasars as a function of \tmax.  Note significant difference for the higher values of \tmax.  See text for discussion.}\label{fig:radio}
\end{center}
\end{figure}

We have formed composite spectra separately for the radio loud and radio quiet AGN, adhering to our previously defined bins in \tmax.  Here radio loud was taken simply to mean detected in the FIRST radio survey  \citep{becker95}.  This is not equivalent to
the standard definition \citep{kellermann89} and is distance-dependent, but it is sufficient to look for any qualitative difference
between radio loud and quiet AGN.  Figure \ref{fig:radio} shows the observed trend of the [\nev] intensity with \tmax\ for the radio loud and quiet composites.
The  [\nev] /[\neiii] ratio is similar for the two groups for $\logtmax = 5.0$, but there is a decline for the radio detected objects with increasing \tmax. 
Certainly, there is no evidence that  [\nev] is stronger for the radio-detected objects, as might be expected if they have higher spin of a prograde
sense.  On the other hand, the weaker  [\nev] of the radio objects might be expected if they involve retrograde disks.

\section{Conclusion}
\label{sec:conclusion}

The results presented here give focus to concerns about the disk model for the AGN continuum that have been voiced in the literature by numerous
authors.   In general terms, AGN over a wide range of \mbh\ and luminosity have rather similar SEDs \citep[see][and references therein]{laor11}.
Organization of the observational data in terms of \tmax\ helps to clarify what trends are expected.  We find that the predicted trends in the intensity
of \hbeta, \heii\ and ]\nev] are not seen in the observations. Over
a factor of five in the value of \tmax\ as inferred from \mbh\ and \nuLnu, the [\nev] intensity shows no systematic variation, and the
\hbeta\ equivalent width shows a reverse trend.  While our models are highly simplified, in particular for cases that approach or exceed the
Eddington limit, the basic immunity of these key line ratios as well as \alphaos\ to the expected value of \tmax\ is a striking empirical fact.
These results suggest that a serious re-examination of the structure and emission mechanisms of the AGN central engine is in order.   This
might reasonably start with more detailed studies of the nature of disks in systems with high Eddington ratios, taking account of advection,
winds, and all contributing sources of opacity.

\acknowledgements
We thank Ivan Hubeny for the use of the AGNSPEC program, 
 R. Antonucci, O. Blaes, A. Laor, and B. Wills  for helpful discussions.   
G.S. gratefully acknowledges the support of the
Jane and Roland Blumberg Centennial Professorship in Astronomy.  
Funding for the Sloan Digital Sky Survey (SDSS) has been provided by
the Alfred P. Sloan Foundation, the Participating Institutions, the
National Aeronautics and Space Administration, the National Science
Foundation, the U.S. Department of Energy, the Japanese
Monbukagakusho, and the Max Planck Society. The SDSS Web site is
http://www.sdss.org/. The SDSS is managed by the Astrophysical
Research Consortium (ARC) for the Participating Institutions. The
Participating Institutions are The University of Chicago, Fermilab,
the Institute for Advanced Study, the Japan Participation Group, The
Johns Hopkins University, the Korean Scientist Group, Los Alamos
National Laboratory, the Max-Planck-Institute for Astronomy (MPIA),
the Max-Planck-Institute for Astrophysics (MPA), New Mexico State
University, University of Pittsburgh, University of Portsmouth,
Princeton University, the United States Naval Observatory, and the
University of Washington.

\end{document}